\PassOptionsToPackage{pdfpagelabels=false}{hyperref} 
\documentclass[usenatbib]{mnras_tex_edited}

\usepackage{newtxtext,newtxmath}

\usepackage[T1]{fontenc}
\usepackage{ae,aecompl}

\usepackage{graphicx}	%
\usepackage{pdfpages}
\usepackage{float}
\usepackage{booktabs}
\usepackage{multirow}
\usepackage{array}
\usepackage{soul}
\usepackage{xcolor}
\definecolor{darkolivegreen}{rgb}{0.33, 0.42, 0.18}

\usepackage{color}

\newcommand{\kms}{km/s}
\newcommand{\msun}{$\rm M_{\odot}$}

\usepackage{pifont}% http://ctan.org/pkg/pifont

\def\aj{\rm{AJ}}
\def\araa{\rm{ARA\&A}}
\def\apj{\rm{ApJ}}
\def\apjl{\rm{ApJ}}
\def\apjs{\rm{ApJS}}
\def\aap{\rm{A\&A}}

\def\mnras{\rm{MNRAS}}
\def\prd{\rm{PRD}}
\def\prl{\rm{PRL}}

\def\nat{\rm{Nature}}

\def\cqg{\rm{CQG}}

\title[Massive black hole triples]{MBH binary intruders: triple systems from cosmological simulations} 
\author[M. Sayeb et al.]{Mohammad Sayeb$^{1}$\thanks{E-mail: \href{mailto:sayebms@gmail.com}{sayebms@gmail.com}},
Laura Blecha$^{1}$,
Luke Zoltan Kelley$^{2}$
\bigskip \\
$^{1}$Department of Physics, University of Florida, Gainesville, FL 32611, USA\\
$^{2}$Center for Interdisciplinary Exploration and Research in Astrophysics (CIERA) and \\\hspace{0.7em}Department of Physics,  University of California, Berkeley, CA 94720, USA\\
}

\pubyear{2020}

\begin{document}
\label{firstpage}
\pagerange{\pageref{firstpage}--\pageref{lastpage}}
\maketitle

\begin{abstract} 

Massive black hole (MBH) binaries  can form following a galaxy merger, but this may not always lead to a MBH binary merger within a Hubble time. The merger timescale depends on how efficiently the MBHs lose orbital energy to the gas and stellar background, and to gravitational waves (GWs). In systems where these mechanisms are inefficient, the binary inspiral time can be long enough for a subsequent galaxy merger to bring a third MBH into the system. In this work, we identify and characterize the population of triple MBH systems in the Illustris cosmological hydrodynamic simulation. We find a substantial occurrence rate of triple MBH systems: in our fiducial model, 22\% of all binary systems form triples, and $>70\%$ of these involve binaries that would not otherwise merge by $z=0$. Furthermore, a significant subset of triples (6\% of all binaries, or more than a quarter of all triples) form a triple system at parsec scales, where the three BHs are most likely to undergo a strong three-body interaction. Crucially, we find that the rate of triple occurrence has only a weak dependence on key parameters of the binary inspiral model (binary eccentricity and stellar loss-cone refilling rate). We also do not observe strong trends in the host galaxy properties for binary versus triple MBH populations. Our results demonstrate the potential for triple systems to increase MBH merger rates, thereby enhancing the low-frequency GW signals detectable with pulsar timing arrays and with LISA.

\hspace{1em}
\end{abstract}

\begin{keywords}
quasars: supermassive black holes -- gravitational waves
\end{keywords}

\section{Introduction}\label{intro}

Many theoretical and observational studies have shown that massive black holes (MBH) reside at the center of most galaxies \citep[e.g.,][]{1998Natur.395A..14R}. There is compelling evidence that the central MBH masses correlate with the luminosity, mass, and velocity dispersion of the galactic stellar bulge \citep{2000ApJ...539L...9F, 2000ApJ...539L..13G, 2002ApJ...574..740T, 2006ApJ...644L..21F}. This indicates a coordinated evolutionary path for MBHs and their host galaxies.

Mergers constitute an important evolutionary step for the galaxies and their MBHs, not least because they lead to the formation of MBH binaries. %As the galaxies merge, the MBHs at their centers, if given the right environment, can  eventually form a binary. 
These binaries are the loudest GW sources %of GW 
in the Universe, with chirp frequencies ranging from $\sim$ mHz for $\sim 10^6$ \msun\ MBHs to $\sim$ nHz for $\sim 10^9$ \msun\ MBHs 
\citep{1994MNRAS.269..199H, 2003ApJ...583..616J, 2003ApJ...595..614W, 2004ApJ...611..623S, 2005ApJ...623...23S}. Recently, Pulsar Timing Array (PTA) experiments around the globe presented strong evidence for a nHz GW background that is consistent with a population of MBH binaries \citep{2023ApJ...951L...8A, 2023arXiv230616214A, 2023ApJ...951L...6R, 2023RAA....23g5024X}.
%These GWs from MBH binaries may soon be detected \citep{2020ApJ...905L..34A, 2021ApJ...911L..34P}. Pulsar Timing Array (PTA) \citep{2010CQGra..27h4013H, 2013CQGra..30v4009K, 2013PASA...30...17M, 2021AAS...23710101D}  experiments are sensitive to nHz GWs (corresponding to $\gtrsim 10^9$\msun\ MBH binaries) and have placed upper limits on the stochastic GW background from unresolved MBH binaries. Recent results indicate that detection of this GW background may be imminent \citep{2020ApJ...905L..34A, 2021ApJ...917L..19G, 2021MNRAS.508.4970C, 2021ApJ...911L..34P}. 
In the coming years, the Laser Interferometer Space Antenna (LISA) will be able to detect mHz GWs from MBH mergers in the $\lesssim 10^6$\msun\ range out to $z\sim20$  
\citep{2017arXiv170200786A}. However, the detection of GWs is only possible if the binary system can reach the GW-dominated regime of inspiral ($\sim$ mpc scales) within a Hubble time. 

The timescales for MBH binaries to inspiral and merge are highly uncertain, which is a major limitation on our ability to predict the GW and EM signatures of the MBH binary population. MBH binary inspiral, also called ``binary hardening", is driven by several processes at different spatial scales \citep{1980Natur.287..307B}. During the galaxy merger, the MBHs fall toward the galactic center via dynamical friction (DF)  \citep{ 1992ApJ...400..460H, 2012ApJ...745...83A}. When the mass enclosed in the binary orbit is comparable to the MBH masses, the binary forms a gravitationally bound system. This typically happens at scales $\lesssim$ tens of pc 
\citep[e.g.,][]{1980Natur.287..307B, 1996NewA....1...35Q, 2002MNRAS.331..935Y}. 

Further hardening of the binary happens via individual stellar scattering events. 
The region of stellar orbital phase space that can interact with the binary is called the loss cone (LC) \citep{1980Natur.287..307B, 1996NewA....1...35Q, 1997NewA....2..533Q, 2013CQGra..30x4005M}. If the loss cone is depleted more quickly than it can be replenished by two-body stellar relaxation, there is no guarantee that the binary can merge within a Hubble time $t_{\rm H}$. This so-called ``final parsec problem" \citep{1980Natur.287..307B, 2003AIPC..686..201M} could be ameliorated by efficient LC refilling in galaxies with triaxial, merging, or otherwise asymmetric potentials \citep{2002MNRAS.331..935Y, 2006astro.ph..1520H, 2006ApJ...642L..21B,    2010ApJ...713.1016H, 2011ApJ...732L..26P, 2011ApJ...732...89K, 2016PhRvD..93d4007K}. Binary hardening can also happen through the interactions with a circumbinary disc in  gas-rich mergers \citep{2007MNRAS.379..956D, 2009MNRAS.393.1423C, 2011MNRAS.412.1591N, 2017MNRAS.472..514G}, although it is unclear whether this mechanism can efficiently drive the binary into the gravitational wave (GW) dominated regime \citep{2009MNRAS.398.1392L, 2019ApJ...871...84M, 2019ApJ...875...66M, 2020ApJ...889..114M} and prevent stalling. Nonetheless, many systems may still experience a significant delay between galaxy coalescence and MBH merger; for example, \citet{2017MNRAS.464.3131K} finds hardening time scales of many Gyr for some systems even with efficient LC refilling.

Triple MBH interactions provide a potential solution to this problem \citep{ 1962AJ.....67..591K, 2018MNRAS.473.3410R, 2018MNRAS.477.3910B}. If the binary inspiral time is longer than the time until the next galaxy merger occurs, a third MBH can enter the system. In cases where this third MBH reaches the host nucleus of the binary before the binary merges, the MBHs can undergo a three-body scattering interaction. Triple interactions can drive stalled binaries to merger on a shorter time scale through Kozai-Lidov (K-L) oscillations in hierarchical triple systems \citep{1962AJ.....67..591K, 1962P&SS....9..719L, 2016ARA&A..54..441N} or through strong three-body encounters. This can potentially increase the merger rates of MBHs, and if binary stalling is common, triple interactions may be essential for driving MBHs to the GW regime where they can be observed by PTAs and LISA.

According to the K-L mechanism, if an intruder MBH forms a hierarchical triple system with the inner binary at a sufficiently inclined orbit, the system will undergo large oscillations in mutual orbital inclination and in the eccentricity of the inner binary. At very high eccentricities, GW emission can drive the MBHs to rapid merger. One caveat to this picture is that such a hierarchical triple will not evolve in isolation, but rather in the stellar and gaseous background of a galactic nucleus. Particularly in gas-rich galaxies, dynamical interactions with this background may dampen the eccentricity oscillations, thereby weakening the effects of the K-L mechanism. However, as noted above, binary hardening via gas dynamical interactions may also prevent MBH binaries from stalling in the first place. In dry systems, where the stalling scenario is more likely, the oscillations induced by K-L mechanism on the inner binary eccentricity could provide an effective means for rapidly driving the inner binary into the GW dominated regime. 

If the intruder MBH reaches the galactic nucleus at a distance comparable to the inner binary separation, a strong, chaotic, three-body interaction between the MBHs will occur. Typically, such an interaction will result in the slingshot ejection of the lightest MBH from the system, while the more massive pair forms a more tightly bound system that can merge on a much shorter timescale \citep{1974ApJ...190..253S, 1975AJ.....80..809H, 2002ApJ...578..775B, 2006ApJ...651.1059I}.

Statistics of close triple MBH encounters were studied in more detail by \citet{2007MNRAS.377..957H}. The authors found that the triple interactions increase the coalescence rate of MBHs and cause spikes of GW emission during the close encounters. A more systematic study of the statistical outcomes of triple MBH interactions was done by \citet{2016MNRAS.461.4419B} and   \citet{2018MNRAS.477.2599B},  where they included all relativistic corrections up to 2.5PN order, which are very important due to the general relativistic (GR) effects inhibiting the K-L mechanism \citep{1997Natur.386..254H}.

Owing to the strong possibility that a triple interaction will result in a slingshot recoil kick to the lightest MBH, these systems are also relevant for their ability to produce offset, wandering MBHs, which may under some circumstances be observable as offset active galactic nuclei (AGN) \citep[e.g.][]{2016ApJ...829...37B}. A candidate slingshot recoil was recently presented by \citet{2023ApJ...946L..50V}, though further observations are needed to confirm its nature. Wandering MBHs can also be produced by GW recoil kicks, which result from the asymmetric emission of GWs during a  
merger between MBHs with unequal masses or spins \citep[][]{1962PhRv..128.2471P, 1973ApJ...183..657B, 2007PhRvL..99d1103L, 2008ApJ...687L..57V, 2008MNRAS.390.1311B, 2011MNRAS.412.2154B}. Constraining the formation of triple MBH systems is therefore important for understanding the relative role of slingshot recoils versus GW recoils in producing a population of wandering BHs, and for determining the impact of such recoil events on the subsequent GW event rate. \citet[][hereafter SB21]{2021MNRAS.501.2531S} recently implemented a model for MBH binary spin evolution along with a binary inspiral model based on that of \citet{2017MNRAS.471.4508K}. One of their key findings was that a non-negligible population of misaligned MBH binaries should exist even if gas dynamical interactions are effective at aligning MBH spins in gas-rich galaxies. Upon merger, these misaligned MBH binaries produce much larger GW recoil kicks, such that the GW recoil distribution always has a high-velocity tail ($v_{\rm kick}>1000$ km s$^{-1}$). 

In order to determine the importance of triple interactions in hastening binary inspiral, producing GW sources, and ejecting MBHs from galactic nuclei, we need to know where and under what circumstances such triple systems actually form. The answer is subject to numerous large uncertainties, not least of which is the uncertainty in the underlying MBH binary inspiral timescales. \citet{2018MNRAS.477.3910B} recently combined their triple interaction modeling results with a semi-analytic model for galaxy evolution in order to predict the role of triple encounters in MBH evolution. They concluded that even if all MBH binaries stall, triple encounters should produce a stochastic GW background that is observable with PTAs. A follow-up study concluded that triple interactions could also contribute to a significant fraction of LISA events \citep{2019MNRAS.486.4044B}.

Models of MBH binary inspiral applied to cosmological hydrodynamics simulations also suggest an important role for triple interactions. Using the Illustris cosmological simulations, \citet{2017MNRAS.464.3131K} find that in their fiducial model, roughly half of all MBH binaries have not merged by $z=0$, and binary inspiral timescales of many Gyr are common. Subsequent studies using variations of this model have found similar results \citep{2017MNRAS.471.4508K, 2020MNRAS.491.2301K, 2021MNRAS.501.2531S}.  \citet{2017MNRAS.464.3131K} estimate that about 30\% of all MBH binaries experience a subsequent merger with another MBH that overtakes the first binary it merges, and in nearly all of these the first binary is overtaken before $z=0$. (Note that these post-processing models evolve the sub-resolution inspiral of all MBH binaries in isolation.)

The primary goal of this work is to examine in much greater detail the properties and characteristics of the triple MBHs and their host galaxies in Illustris. This constitutes the first dedicated study of triple MBH formation based on a cosmological hydrodynamics simulation. We use the same hardening prescription as in SB21 for a population of merging binaries  \citep[based on][]{2017MNRAS.471.4508K} from Illustris. We identify cases where an intruder takes over a binary and forms a triple system. We look at characteristics of these systems, their environments and their host properties. We also study the dependence of these triples on variations of model parameters. Our results provide a new type of cosmological framework for understanding the role of triple interactions in MBH evolution and GW source populations, including their environments and host galaxy properties.

In Section \ref{methods} we describe the MBH population, our hardening prescription, and the method that we use for the identification of the triples. In Section \ref{results} we describe our findings, and finally in Section \ref{discussion} we discuss the caveats of our finding, and draw conclusions.

\section{Methods}\label{methods}

Our study focuses on the population of merging MBHs from the Illustris cosmological hydrodynamic simulation suite\footnote{\url{http://www.illustris-project.org/}} \citep{2014MNRAS.444.1518V, 2014MNRAS.445..175G, {2015A&C....13...12N}}. Due to the large scale of these simulations and the computational limitations, a gravitational softening scale is defined for each particle or cell, below which the simulation is not able to resolve the physics. The MBH mergers in Illustris happen when two MBH particles are within a softening length of each other, which is typically of the order of $\sim$kpc. In reality, however, the MBHs are far from merger at this point. To model binary inspirals  at sub-resolution scales, we implement a prescription that extrapolates the central density profiles of host galaxies from Illustris to calculate hardening rates due to various mechanisms: dynamical friction, stellar scattering, circumbinary gas disk-driven hardening, and GW emission \citep{2017MNRAS.464.3131K, 2017MNRAS.471.4508K, 2021MNRAS.501.2531S}. To identify triples, we find MBHs that experience more than one merger and track each merger in redshift and MBH separation space ($z,\;a(z)$) to find if a merger is overtaken by a subsequent merger (i.e. if the $a_{1}(z)$ and $a_{2}(z)$ curves of the two mergers cross; Figure \ref{triple_tree}). 

The binary and triple parameter definitions are as follows: $m_1$ and $m_2$ stand for the masses of the primary MBH and the secondary MBH in the inner binary. The primary is more massive than the secondary ($m_1>m_2$). The mass ratio of the inner binary is defined as $q_{\rm inner}=m_2/m_1$, which makes it by definition smaller than unity. The combined mass of the inner binary is defined as $M_{\rm bin}=m_1+m_2$. As for the outer binary, the mass of the intruder is indicated by $m_3$. For the outer binary mass ratio $q_{\rm outer}$, however, we want to distinguish between the ``intruder" MBH and the MBH that is a member of the inner binary. Thus, the outer binary mass ratio is determined by:

\begin{equation}
    \label{mass_ratio_definition}
    q_{\rm outer} = \frac{m_3}{M_{\rm bin}}
\end{equation}

The combined mass of the binary and the intruder are defined as $M_{\rm tot} = m_1+m_2+m_3$.

\subsection{Illustris}\label{illustris_sim}

The Illustris simulations are a suite of cosmological hydrodynamic simulations run using the \textsc{arepo} hydrodynamics code \citep{2010MNRAS.401..791S}. The  \textsc{arepo} code is based on an unstructured moving mesh that combines the advantages of smooth particle hydrodynamics (SPH) \citep[e.g.][]{1977MNRAS.181..375G, 1977AJ.....82.1013L} and an Eulerian mesh-based approach \citep[e.g.][]{1989JCoPh..82...64B}. The mesh is formed from  Voronoi tessellations based on a set of discrete mesh-generating seeds that can freely move and create a dynamic topology  \citep{2010MNRAS.401..791S}. Dark matter (DM), star, and MBH particles are superposed on the mesh, where MBHs of mass $1.4\times10^5$\msun\ are seeded in the center of halos with total mass $>7.1\times10^{10}$\msun\ that do not already have a MBH \citep{2015MNRAS.452..575S}. The MBHs can merge, accrete gas, and impart AGN feedback energy and momentum to their surroundings. \citet{2013MNRAS.436.3031V} describes the details of the sub-grid MBH accretion and feedback prescriptions, along with the sub-grid prescriptions for other physics including star formation, stellar feedback, chemical enrichment, and metal-line cooling. The Illustris simulation  reproduces reasonably well many observed properties of galaxies and their MBHs, including the galaxy stellar mass function, stellar luminosity functions, cosmic star formation history, baryon conversion efficiency, the MBH mass function, and AGN luminosity functions \citep[e.g.,][] {2014MNRAS.444.1518V,  2014MNRAS.445..175G, 2015MNRAS.452..575S}.

Galaxies in Illustris are identified via the Subfind algorithm \citep{2009MNRAS.399..497D}, which finds gravitationally bound ``subhalos" within friends-of-friends halos. Throughout this paper, any mention of MBH host galaxies in Illustris refers to the Subfind subhalos that host the MBHs.

The Illustris simulations are run in a cosmological box of side $L_{\rm box}=75\, h^{-1}$ Mpc, from $z=137$ to $z=0$. Throughout this paper we use the highest-resolution run, `Illustris-1' (hereafter `Illustris'), which has a DM resolution of $6.3\times10^6$\msun\ and a baryonic mass resolution of $1.2\times10^6$\msun. The simulations assume a WMAP9 cosmology with parameters $\Omega_{m}=0.2865$, $\Omega_{\Lambda}=0.7135$, $\sigma_8=0.820$, and $H_0=70.4$ \kms\ Mpc$^{-1}$ \citep{2013ApJS..208...19H}.

As in SB21, we use the Illustris simulation for our analysis rather than the more recent IllustrisTNG simulations\footnote{\url{https://www.tng-project.org}} \citep{ 2018MNRAS.475..624N, 2018MNRAS.475..676S, 2018MNRAS.480.5113M, 2018MNRAS.475..648P, 2018MNRAS.477.1206N, 2019MNRAS.490.3234N, 2019MNRAS.490.3196P} because this enables us to use the existing MBH binary inspiral data from \citet{2016MNRAS.456..961B, 2017MNRAS.464.3131K}. These inspiral models were obtained by extracting stellar, gas, and DM density profiles for the hosts of all merging MBHs and integrating the hardening rates from the initial to final separation for each binary. By using Illustris for this study, we are also able to compare directly with results of previous binary inspiral studies based on this simulation \citep[][SB21]{2016MNRAS.456..961B, 2017MNRAS.464.3131K, 2018MNRAS.477..964K, 2020MNRAS.491.2301K}. Although IllustrisTNG has been shown to better reproduce numerous stellar and gas properties of galaxies \citep{2018MNRAS.473.4077P, 2018MNRAS.475..648P}, both Illustris and IllustrisTNG produce MBH populations that broadly agree with empirical constraints \citep[][]{2015MNRAS.452..575S, 2018MNRAS.473.4077P, 2018MNRAS.475..648P, 2021MNRAS.503.1940H}. Aside from this the SMBH seeding in TNG is higher than the Illustris by a factor of $\approx 8$ \citep{2014Natur.509..177V, 2018MNRAS.473.4077P}. This, with our prescription for SMBH population delineated in the next section, would exclude more SMBHs within LISA sensitivity range. The IllustrisTNG simulation suite does, however, include both higher-resolution (50 Mpc)$^3$ and lower-resolution (300 Mpc)$^3$ volumes in addition to the fiducial (100 Mpc)$^3$ volume, which would enable more detailed studies of the inner structure of galaxies and the highest-mass MBH populations, respectively. In future work, we plan to generalize our analysis to other datasets.

\subsection{Massive Black Hole Binary Population}\label{mbh_population}

The time at which Illustris records an MBH merger is the starting point for our post-processing binary evolution model. Thus, we refer to the Illustris merger time as the binary {\em formation} time. The MBH binary population studied here is identical to that of SB21. Both in this paper and in SB21 we include a subset of all 23708 MBH mergers in the Illustris for our study. Illustris uses an MBH re-positioning scheme that places MBHs at the potential minimum of their host halos. This is done in order to prevent spurious numerical kicks to the MBHs by nearby particles or cells, which have comparable or even greater mass. However, in some unequal mass mergers, a  satellite galaxy can lose its MBH to the more massive galaxy on an artificially short time scale, and in some cases the satellite can then be re-seeded with a new MBH. This process can happen multiple times and create spurious mergers, particularly for MBHs near the seed mass. In order to mitigate the impact of these issues on our results, we follow previous work and exclude all mergers that have MBHs below $10^6$\msun\ \citep[][SB21]{2016MNRAS.456..961B, 2017MNRAS.464.3131K}. We note that the excluded range of MBH masses ($10^5$ - $10^6$ \msun) aligns with the peak sensitivity of LISA to MBH merger events. However, the simple, massive MBH seeding prescriptions in Illustris and other large-volume cosmological simulations limits their predictive ability for LISA in any case \citep[e.g.,][SB21]{2016MNRAS.463..870S, 2020MNRAS.491.2301K}.

The binary hardening prescription in \citet{2017MNRAS.464.3131K} that we are using requires the binary to have an associated galaxy in the snapshots before and after the merger. This is required to identify MBH hardening environments to calculate dynamical friction, stellar scattering, and disk hardening rates. There are further constraints on the number of each particle type in a galaxy for it to be resolvable. We require a minimum of 80 gas cells, 80 star particles, and 300 DM particles for a galaxy to be resolved.   A combination of all these constraints reduces the total binary population that we are studying to 9234 systems. 

\begin{figure*}
    \centering
    \includegraphics[width=0.37\textwidth, trim=0 -2cm 4cm 0, clip]{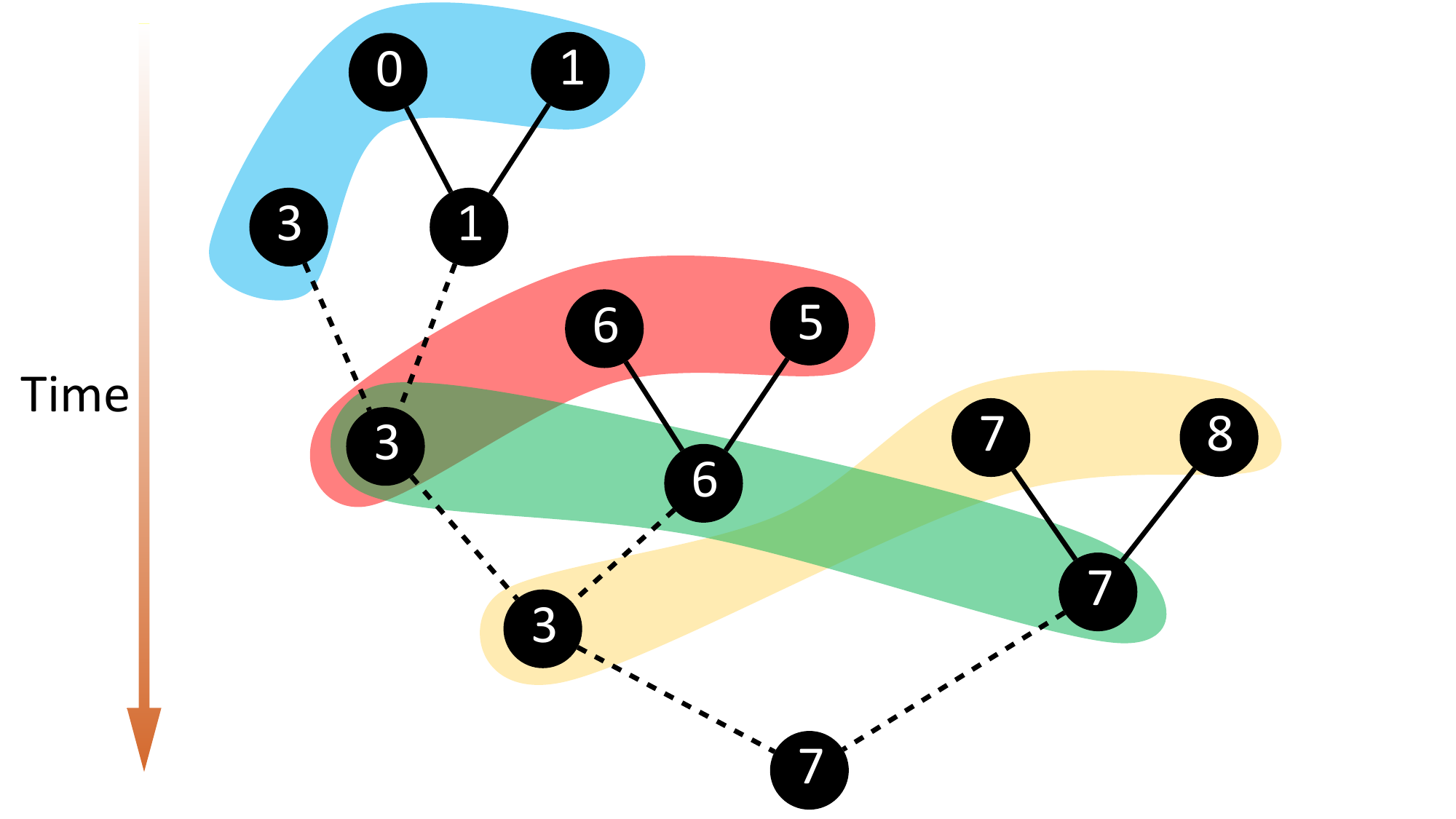}
    \includegraphics[width=0.62\textwidth, trim=1cm 4cm 1.5cm 4cm, clip]{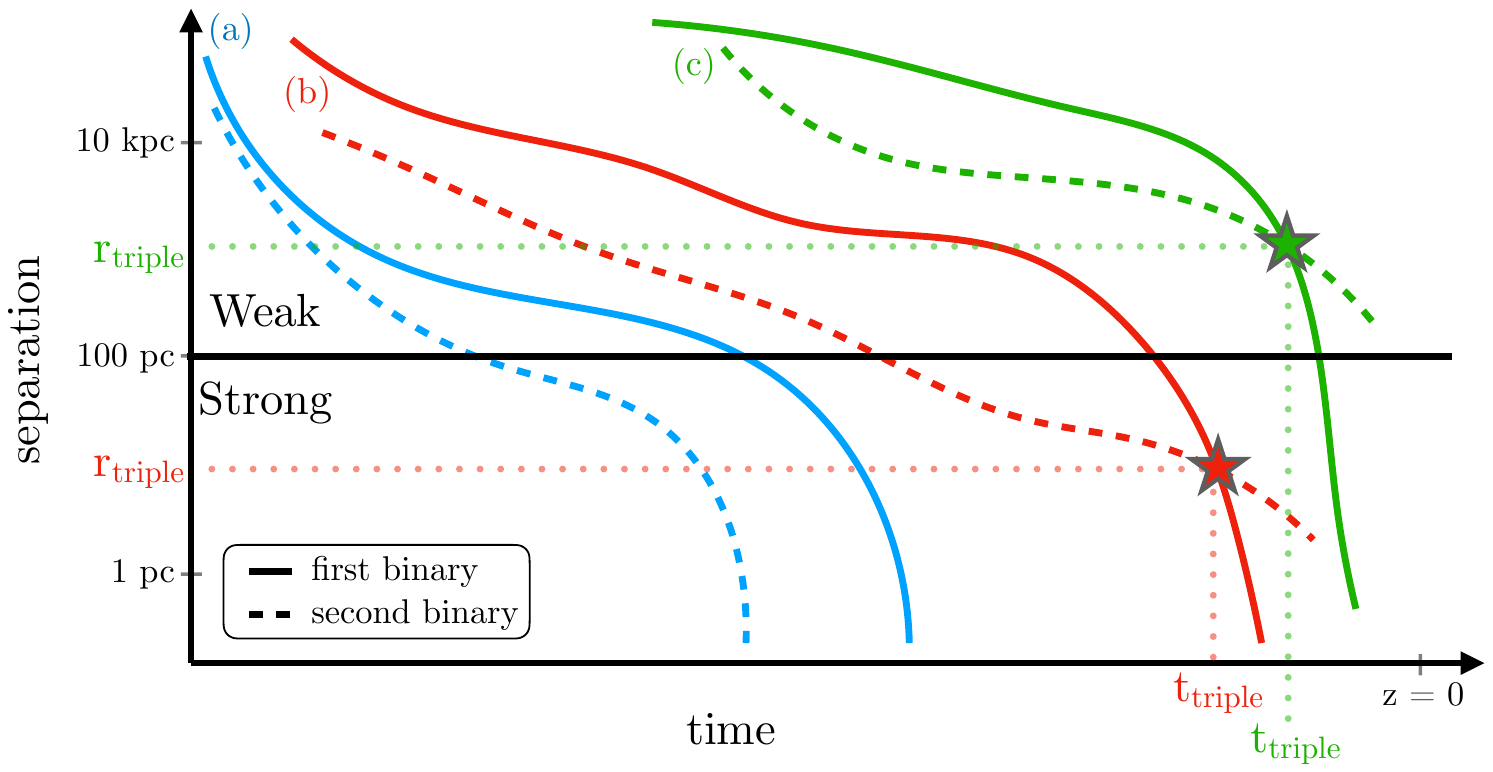}%
    \caption[Triple merger tree and triple radii]{ A schematic of MBH merger tree (left) and separation vs time evolution (right) are shown here. The dark circles represent MBHs in the left hand side. The numbers in each circle represent the MBHs unique ID in the Illustris simulation and the direction of time flow is from top to bottom. The highlighted areas show potential independent triples. Note that the color of the highlighted area on the merger tree and the schematic curves plotted on the right hand side are arbitrary; there is no one-to-one correspondence between the merger tree on the left and the schematic curve on the right. However, the solid and dashed linking lines in the diagram on the left do indicate the first and second binaries, respectively.  On the right hand side the curves show the binary separation as a function of time. The solid curves show the first merger and the dashed lines show the subsequent or second merger. In order to identify the systems that can potentially evolve into a triple MBH, we look for repeated MBH IDs in subsequent mergers (left). If a repeated ID is identified we then look at the separation vs time curve of the current binary (i.e. first binary or inner binary) and that of the subsequent binary (i.e. second binary or outer binary). If the curves cross at some point (denoted with stars) before $z=0$ then the system becomes a successful triple. Three different classes of outcomes are shown in the right figure. (a) The blue curves do not intersect with each other, meaning that the first binary is being chased by the intruding MBH (outer binary) and no triple forms. (b) The red curves show a case where the first binary is overtaken by the intruder (outer binary) at small separations ($< 100$ pc); we refer to this as a ``strong triple". (c) The green curves show a binary overtaken by an intruder at large separations, indicating a ``weak triple".}
    \label{triple_tree}
\end{figure*}

\subsection{Hardening Prescription}\label{hardening_prescription}

We follow the same hardening prescription as in SB21. This prescription is outlined in \citet{2017MNRAS.464.3131K, 2017MNRAS.471.4508K} and implements the four different hardening mechanisms \citep{1980Natur.287..307B} of dynamical friction (DF), stellar scattering also knows as loss cone scattering (LC), circumbinary disk hardening (CD), and hardening due to emission of gravitational waves (GW). Hardening rates for each are calculated using inward extrapolations of spherically averaged stellar, DM, and gas density profiles. We will briefly go over them and how they are calculated here; we refer the reader to \citet{2017MNRAS.464.3131K}, \citet{2017MNRAS.471.4508K}, and SB21 for more details.

\begin{figure*}
    \centering
    \includegraphics[width=\textwidth, trim=1cm 0.5cm 0.5cm 0, clip]{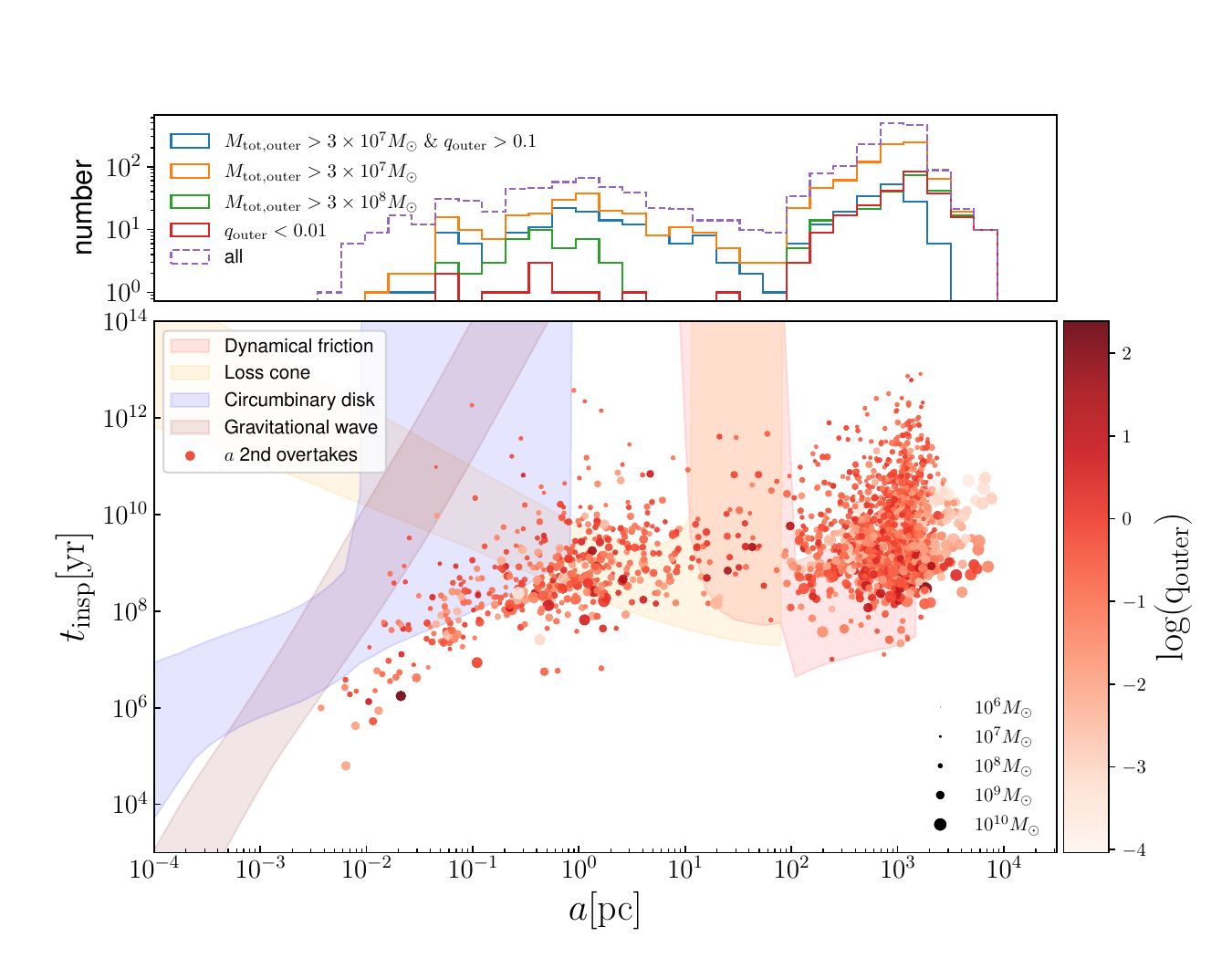}
    \caption[Triple inspiral time and radii]{
    The shaded regions in this plot show the inspiral time scale vs.~separation for the MBH binaries in our fiducial model, with each color corresponding to a different hardening mechanism as shown in the legend (from SB21 Figure 2). The scatter plot  shows $a_{\rm triple}$ and the corresponding total inspiral time for the subset of binaries that become triples (specifically, these data correspond to the outer binaries in each triple system). The scatter points are color-scaled to the logarithm of the mass ratio $q_{\rm outer}$ of the outer binary and the size of each point correspond to the total mass of the outer binary. The top histogram shows the distribution of $a_{\rm triple}$ for the triple population for five different sub-populations: all triples, triples with $\rm q_{\rm outer}<0.1$, $M_{\rm tot}>3\times10^8$\msun, $M_{\rm tot}>3\times10^7$\msun, and  $M_{\rm tot}>3\times10^7$\msun \:\& \:$q_{\rm outer}>0.1$, and triples with $M_{\rm tot}>3\times10^8M_{\odot}$ Notice the bi-modality in the $a_{\rm triple}$ distribution. This bi-modality constitutes the basis of our definition of strong and weak triple interactions.}
    \label{fig:hardening_scatter}
\end{figure*}

At large binary separations ($\sim$kpc), dynamical friction is the only effective binary hardening mechanism (Figure~\ref{fig:hardening_scatter}, shaded regions) \citep[e.g.,][]{2012ApJ...745...83A, 2017MNRAS.464.3131K}. At separations of $\sim$100 pc--- $\sim$0.1 pc, LC stellar scattering drives the hardening of the binary. In this paper, we consider two implementations of hardening in the LC regime:

\begin{itemize}
\item {\bf Variable  
LC refilling rate, zero eccentricity}\\
This model allows for systematic variation of the loss-cone refilling efficiency. For variations of the loss cone refill fraction in our model, we choose circular orbits ($e=0$) for all the binaries. The loss-cone refilling efficiency is based on dynamical models of stellar scattering rates from \cite{1999MNRAS.309..447M}. 
In a `full' LC, low-angular-momentum stars are replenished as fast as they are scattered out of the LC. If instead, two-body relaxation is the only mechanism by which the LC is replenished (on timescales $\gg t_{\rm H}$) then we have a `steady-state' loss-cone. Our models consider these two scenarios for circular orbits, following \citet{1999MNRAS.309..447M}, as well as intermediate scenarios (a partially full LC). We use the logarithmic `refilling fraction' $\mathcal{F_{\rm refill}}$, to interpolate between the stellar flux for a full LC ($F^{\rm full}_{\rm LC}$) and that of a steady-state LC ($F^{\rm eq}_{\rm LC}$):

\begin{equation}
    F_{\rm LC}\equiv F^{\rm eq}_{\rm LC}\times \Bigg(\frac{F^{\rm full}_{\rm LC}}{F^{\rm eq}_{\rm LC}}\Bigg)^{\mathcal{F_{\rm refill}}}.
    \label{refill_fraction}
\end{equation}

Following SB21, we adopt $\mathcal{F_{\rm refill}}=0.6$ as the fiducial value, corresponding to a partially full LC and we compare to results for $\mathcal{F_{\rm refill}}=0$ \& $1$ in Section {\ref{model_dependence}}.

\item {\bf Variable binary eccentricity, full LC} \\
This model allows for the eccentricity of the binary to evolve over time and is based on the scattering experiments by \cite{2006ApJ...651..392S}. The LC hardening rate $da/dt$ and eccentricity evolution $de/dt$ are:

\begin{equation}
\left (\frac{da}{dt} \right )_{\rm LC} \equiv -\frac{G\rho}{\sigma}a^2 H
    \label{dadt_lc}
\end{equation}

\begin{equation}
\left ( \frac{de}{dt} \right )_{\rm LC} \equiv \frac{G\rho}{\sigma}a HK
    \label{dedt_lc}
\end{equation}

Here $a$ is the binary separation $e$ is the binary eccentricity. $\rho$ and $\sigma$ are stellar  density and stellar velocity dispersion profiles calculated from the corresponding host in the Illustris simulation. $H$ and $K$ are dimensionless constants determined by numerical scattering experiments \citep[we use the values from][]{2006ApJ...651..392S}. In all realizations of the binary population using this model, the initial eccentricity is set to the same value for all binaries; we consider models with initial eccentrities of 0.1 to 0.99. 

The GW hardening rate and eccentricity evolution are \citep{1964PhRv..136.1224P}:

\begin{equation}
 \left(\frac{d a}{dt}\right)_{\rm GW}=-\frac{64G^3}{5c^5} \frac{m_1 m_2 \left	(m_1 +m_2\right)}{a^3} \frac{(1+ 73 e^2/24+ 37e^4/96)}
 {(1-e^2)^{7/2}}.
 \label{dadt_gw}
\end{equation}

\begin{equation}
    \left(\frac{d e}{dt}\right)_{\rm GW}=-\frac{304G^3}{15c^5} \frac{m_1 m_2 \left	(m_1 +m_2\right)}{a^4} \frac{(e+ \frac{121}{304}e^3)}
 {(1-e^2)^{5/2}}
 \label{dedt_gw}
\end{equation}

Note that in the LC phase the eccentricity increases as the binary hardens, while in the GW phase binary rapidly circularizes. Note also that eccentricity does not evolve in the disk phase (CD) in our model. 

\end{itemize}

\begin{figure*}
    \centering
    \begin{minipage}[b]{.48\textwidth}
    \includegraphics[width=3.55in, trim=0.2in 0 0.2in 0, clip]{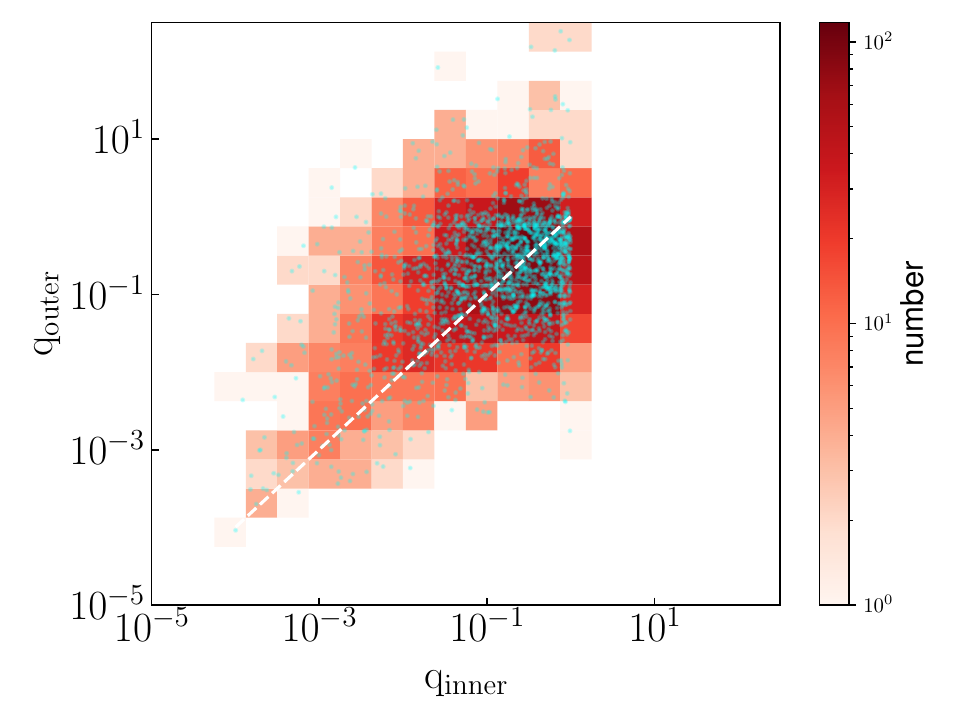}
    \end{minipage}\qquad
    \begin{minipage}[b]{.48\textwidth}
    \includegraphics[width=3.55in, trim=0.2in 0 0.2in 0, clip]{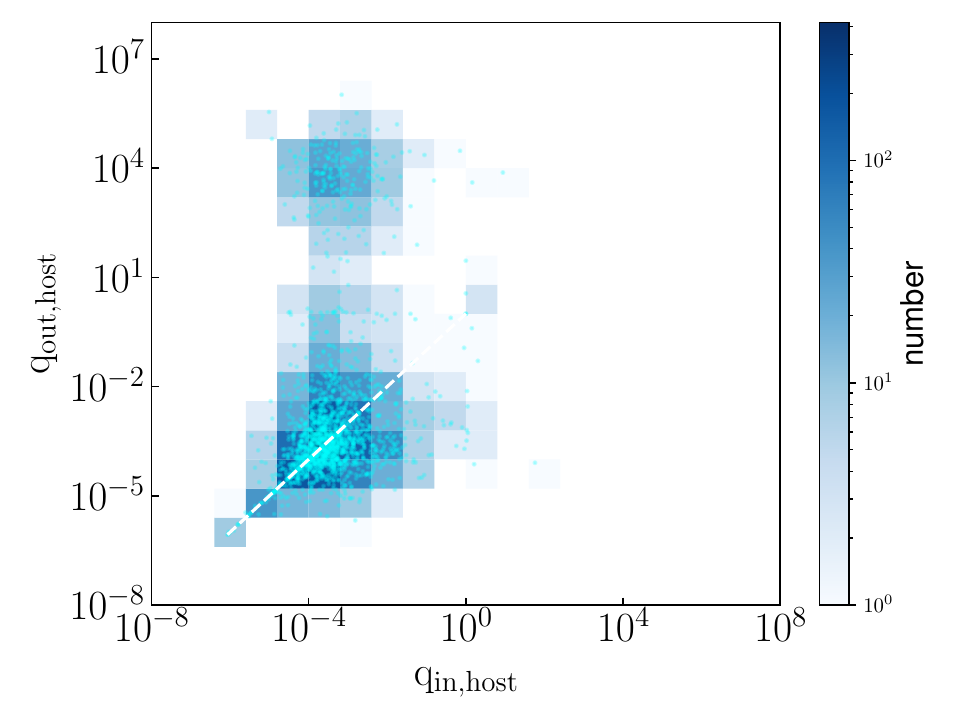}
    \end{minipage}
    \caption[MBH and subhalo mass ratios]{Mass ratio of the outer binary MBH (y-axis) and the inner binary MBH (x-axis) is shown in the left here. On the right we have the mass ratio of the corresponding subhalos. The white dashed lines indicates the one to-one ratio. We see that the ${\rm q_{outer}}$ are slightly larger compared to ${\rm q_{outer}}$ for the MBHs. For the subhalo mass ratios we see a similar trend, except that a subset of intruder MBH hosts are very massive relative to the inner binary hosts. 
    }
    \label{qhist_2d}
\end{figure*}

\subsection{Identifying Triples}\label{identifying_triples}

Because our post-processing binary inspiral model evolves each MBH binary in isolation, we must develop a method for identifying which binary systems involving a common MBH coexist in time, and which of these coexisting systems are likely to become true MBH triple systems. The left panel in Figure \ref{triple_tree} shows a schematic of a merger tree from Illustris and a schematic separation versus time diagram for the potential triple systems.  The highlighted regions indicate potential triples.  In order to identify and classify triple systems, we first create a time-ordered array of IDs from the Illustris merger tree. When two MBHs merge in Illustris, one of the MBH IDs survives the merger and is assigned to the remnant MBH. We can see this depicted in Figure \ref{triple_tree}. When for instance MBHs with IDs `00' and `01' merge, the remnant is assigned the ID `01'. Afterwards, we look at repeated MBH IDs in the merger array. Consecutive mergers involving the same MBH ID with a nonzero overlap in their binary inspiral times are referred to as ``coexisting" binaries (recall that the time of the MBH merger in Illustris is the binary formation time in our model). This sample of coexisting binary systems forms the super-set of binaries that have the potential to become triple systems. 

In some coexisting binary systems, the second binary never overtakes the first. Curve $(a)$ in Figure \ref{triple_tree} shows an example of this case, in which no intersection point ($a_{\rm triple}$) is defined. We refer to these systems as ``failed triples'' (FT). Finally, because we are interested in the population of triple MBH systems in the observable Universe, we further restrict the definition of ``triples" to include only those in which the first binary is overtaken by the second {\em before} $z=0$. Thus, binaries that are overtaken at later times $z<0$ are also included in our definition of ``failed triples".

To determine which of these coexisting binaries are most likely to undergo triple interactions, we compare the evolution of binary separation vs time ($a,t$) for each set of coexisting binaries. For a given set of two coexisting binaries involving a common MBH, we identify the ``first" or ``inner" binary as the one with the earlier formation time, and the ``second" or ``outer" binary as the one with the later formation time. The point in time at which the separation of the second binary becomes smaller than the separation of the first binary is defined as the time of formation of the triple ($t_{\rm triple}$). The corresponding binary separation is called the triple formation radius $a_{\rm triple}$. 

We define ``triples" any such coexisting binaries in which the first is overtaken by the second before $z=0$, such that $a_{\rm triple}$ is a defined quantity. The schematic curves $(b)$ and $(c)$ in Figure \ref{triple_tree} illustrate two examples of triple formation at small and large binary separations, respectively. We refer to case (b) as a ``strong triple" (ST) and case (c) as a ``weak triple" (WT), where 100 pc is the critical separation between the two scenarios. Strong triples are of particular interest, as these are the systems in which all three MBHs are most likely to be in close proximity and undergo a dynamical interaction. The motivation for choosing 100 pc for this definition is discussed further in Section \ref{triple_mbh}. 

We note that these binaries are evolved in isolation in a static background density profile in our models, which is a significant simplifying assumption. For example, we cannot account for situations where a binary is not technically overtaken by a second binary, but where the third MBH may be close enough to have some dynamical influence on the first binary. We also cannot account for merger-induced disturbances to the host galaxy potential that may alter the inspiral timescales of coexisting binaries. Our primary aim in this study is to characterize the binary systems most likely to become triple systems, and to identify the subset most likely to have strong triple interactions; a detailed investigation of triple dynamics in time-varying galaxy merger potentials is beyond the scope of this work. However, as discussed below, our findings suggest that these simplifications have a fairly minimal impact on the proportion of binaries that become strongly interacting triple systems.

\section{Results}\label{results}

\subsection{Characteristics of Triple MBH Systems}\label{triple_mbh}

We find that triple MBH systems represent a substantial  portion of the binary population. 35\% of all binaries in our fiducial model coexist with a second  binary at some point during their inspiral (the super-set of all coexisting systems defined above). Most of these coexisting systems (22\% of the total  binary population) go on to become true triple systems, in which the first binary is overtaken by the second binary before $z=0$.  As expected, these triple systems disproportionately occur when the first binary is stalled. In SB21, we found that 53\% of the binaries in our fiducial model  are ``unmerged" or ``stalled"---that is, they do not merge before $z=0$. Based on this fiducial model, 71\%  of triple interactions involve binaries that would not otherwise merge, demonstrating a significant potential for triple systems to drive these unmerged binaries to merger. The remaining 29\% of triple interactions involve binaries that {\em would} merge before $z=0$ according to our fiducial model, but may instead have a different outcome when triple dynamics are considered (for example, a hastened merger, or a merger between the intruder MBH and a member of the first binary). The proportion of unmerged and merged binaries from the fiducial model that end up being triple systems is 30\% and 13\%, respectively. These triple systems, if successful at accelerating the inner binary hardening through either K-L oscillations or three-body scattering, could drive some of the unmerged binaries to merger before $z=0$ and increase the total merger rate.
  
Figure \ref{fig:hardening_scatter} shows a scatter plot of total inspiral timescale for triples at the triple formation radius $a=a_{\rm triple}$ (i.e., the radius at which the second binary overtakes the first, according to our binary inspiral models). These data are overplotted on binary inspiral timescales versus separation for different hardening mechanisms, for all binaries (coexisting or not). Looking at the scatter plot, we can see that the triple formation radii exhibit a strong bi-modality: there are distinct populations of triples that occur at either larger separations ($\sim$ kpc) or smaller separations ($\sim$ pc). As noted in \S\ \ref{identifying_triples}, we adopt a cutoff of 100 pc between these two populations, and we designate the former (overtaken at $>100$ pc separations, with a median $a_{\rm triple}= 992$ pc) as ``weak triple" (WT) systems. The latter (overtaken at $<100$ pc) are designated ``strong triple" (ST) systems ---these are the triple systems most likely to undergo strong three-body interactions. Most triples ($\sim$74\%) form at large separations of $a_{\rm triple} \gtrsim$ 100 pc. The ST systems, in contrast, have a median formation radius of $a_{\rm triple}= 0.9$ pc.

The choice of the 100 pc cutoff between WT and ST formation radii is validated by its simplicity and its statistical consistency. We performed K-means clustering \citep{Macqueen67somemethods} on the data in Figure \ref{fig:hardening_scatter} with features ($q,\; M_{\rm total},\; a$) and achieved very similar results to a simple 100 pc cutoff. The largest drop in the K-means inertia, which is an indicator of the compactness of the cluster (smaller inertia indicates a more compact cluster and vice versa), happens for two clusters, and the boundary between the clusters corresponds to $\approx$ 100pc. The median $a_{\rm triple}$ for WT and STs from the clustering analysis is 980pc and 0.7 pc, respectively; these values are essentially identical to those obtained above. Therefore, we use the 100 pc threshold for ST and WT classifications henceforth.

In our fiducial model, STs and WTs constitute 6\% and 16\% of the total binary population and 16\% and 47\% of the coexisting binary population, respectively; an additional 13\% of all binaries are classified as FTs and are discussed further in \S\ \ref{failed_triples}. Because the intruder MBH in STs must evolve to much smaller separations before overtaking the inner binary, STs have a much longer delay between the formation of the second binary and the time when the intruder MBH overtakes the first binary: the median time for the intruder MBH in STs to overtake the inner binary is $2.8\times10^9$ yr. In contrast, WTs form with a vastly shorter median delay time of $6.3\times 10^7$ yr. The corresponding median formation redshifts for STs and WTs are $z=0.7$ and 1.1, respectively.

At the time of formation of the triple, the median total mass of inner binaries  is $6.5\times10^7M_{\odot}$, while the intruders have a median mass of $1.0\times10^7M_{\odot}$. The inner binaries   have time to accrete gas between the time of binary formation and the time of triple formation and the quoted inner binary mass takes that into account. The median binary masses for the inner binary are similar for weak and strong triples: $\rm M_{\rm bin}=7.0\times10^7$ \msun\ and $\rm M_{\rm bin}=5.8\times10^7$ \msun, respectively. Comparatively the median binary mass for the total binary population, without taking into account the accretion from the time of the binary formation until the time of triple formation for triple MBHs, is $M_{\rm bin}=3.0\times10^7$ \msun. In addition, a significant number of the   binaries (58\% of inner binaries and 64\% of outer binaries) are major mergers with ${q \geq 0.10}$, especially those that form STs. The median inner binary mass ratios are ${q_{\rm inner}=0.10}$ for WTs and ${q_{\rm inner}=0.33}$ for STs. For the outer binary, the mass ratios ($q_{\rm outer}$) are 0.19 and 0.21 for weak and strong triples, respectively.

The size of the points in Figure \ref{fig:hardening_scatter} corresponds to the total mass of the outer binary, while the color scale of the points corresponds to the logarithm of the outer binary mass ratio. Additionally, the top histograms in Figure \ref{fig:hardening_scatter} show the distribution of $a_{\rm triple}$ for all triples, as well as the subsets in which the outer binary is a major merger (with mass ratio $q_{\rm outer}>0.1$) or has a large total mass of $M_{\rm tot}>3\times10^8$\msun. We see that most of the triple systems in which the first binary is a minor merger $q<0.1$ with a high total binary mass tend to form at large separations. Both the WT and ST subsets span a wide range of inner binary mass ratios and total masses; however, nearly all of the extremely massive systems are in the WT regime and have very low mass ratios. This increases the dynamical friction timescales and leads to a larger triple formation radius. In such cases, if the intruder MBH is also much less massive than the central MBH, the WT may consist of two satellite MBHs orbiting in the massive host halo for more than a Hubble time. Conversely, if the intruder MBH is comparable to or more massive than the central MBH ($q_{\rm outer} \gtrsim 1$), its evolution to form a binary with the central MBH will likely be unhindered by the presence of the much less massive, wide-separation member of the first binary.

\begin{table*}
\centering
\begin{tabular}{l|ccccc}
(Sub-)Category & \% of & \% of & \% of & \% with & \% with  \\ 
 & all binaries & coexisting binaries & triples or FTs & $q_{\rm in} \geq 0.1$ & $q_{\rm out} \geq 0.1$\\ \hline
Triples & 22 & 63 & 100  & 58 & 64\\
\hspace{12pt}STs & 6 & 16 & 74 & 21 & 17 \\
\hspace{12pt}WTs & 16 & 47 & 26 & 37 & 47 \\ \hline
FTs & 13 & 37 & 100 & - & - \\
\hspace{12pt}FTs: near-miss STs & 1.3 & 3.7 & 10 & - & - \\
\hspace{12pt}FTs: near-miss WTs & 3.8 & 11 & 30 & - & - \\
\end{tabular}
\caption{Summary of the categorization of all coexisting binaries, as defined in the text. Coexisting binaries are classified as either triples or failed triples (FTs). Triples are further categorized into strong triples (STs) and weak triples (WTs). For FTs, we also show the subsets that nearly miss forming a ST or a WT. For each of these two categories and four subcategories, we report the percentage of all binaries and the percentage of all coexisting binaries that fall in that (sub-)category. We also report the percentage of triples or FTs in each sub-category, and the percentage of each triples (sub-)category for which the inner or outer binaries have a mass ratio $q\geq0.1$.}
\end{table*}

Figure \ref{qhist_2d} shows a 2D histogram of the mass ratio of the outer binary (${q_{\rm outer}}$) versus the inner binary (${q_{\rm inner}}$) for all triple MBHs on the left hand side and the corresponding total subhalo mass ratios of their hosts on the right hand side. The ${q_{\rm inner}}$ and ${q_{\rm outer}}$ mass ratio distributions are fairly similar, but with a large scatter in ${q_{\rm inner}}$ versus ${q_{\rm outer}}$ for individual systems. On average, though, ${q_{\rm outer}}$ in each triple system tends to be slightly higher than ${q_{\rm inner}}$. This may partly reflect the fact that the intruder MBH interacts with the system at a later time and thus has more time to accrete before it enters the system. Another contributing factor can be the fact that inner binaries with low mass ratios (i.e. small ${q_{\rm inner}}$) in general have longer merger timescales, and thus are more likely to be overtaken than inner binaries with equal mass ratios \citep[][SB21]{2017MNRAS.464.3131K}.

We also examine the mass ratios of the host galaxies. We define ${q_{\rm in,host}}$ and $q_{\rm out,host}$ in the same manner as $q_{\rm inner}$ and $q_{\rm outer}$, but using the total subhalo masses of the hosts instead of the MBH masses. These host mass ratios largely show a similar trend to the MBH mass ratios, in which the outer binaries in general have higher mass ratios compared to inner binaries. The host mass ratio plot also shows a bi-modality in the distribution of outer host mass ratios ${q_{\rm out,host}}$ compared to ${q_{\rm in,host}}$. The very high mass ratio systems with ${q_{\rm out,host}}>10$ constitute $\sim 10\%$ of the triple population. Most ($\sim 70\%$) of these very high mass ratio systems are weak triple systems.

\subsection{Failed Triple Systems}
\label{failed_triples}

We now examine in more detail the population of FTs---the $\sim 1/3$ of coexisting MBH binaries that fail to become triples according to the above definitions. The failed triple MBHs (1170 systems, or 13\% of all binaries) consist of two sub-populations. Most of them (79\%) are systems in which the inner binary merges before it can be overtaken by the outer binary. In the rest of the FT systems, the outer binary does eventually overtake the inner binary, but not before $z=0$. Focusing on the former category, we consider whether some of these FTs may have come close to forming a successful triple. Given that our models include only binary evolution in isolation, whereas triple dynamics will be more complicated, it is possible that some such ``near misses" could in reality become triple systems. 

We compare the minimum ratio of the outer binary and the inner binary separations, $(a_2/a_1)_{\rm min}$, in order to see how close the outer binary came to overtaking the inner. Of greatest interest are the systems in this category that have a small binary separation ratio ($1 < (a_2/a_1)_{\rm min} \leq 10$) that {\em also} have a close inner binary at the point of closest approach ($a_1 < 100$ pc). These are the FTs that came closest to forming STs without the intruder actually overtaking the inner binary. Only 120 binaries are in this category (10\% of all FTs, or 1.3\% of all binaries). Given that 6\% of all binaries do form STs in our fiducial model, these near misses would at best provide a small but not insignificant enhancement to the ST population, if most of them were actually able to form successful triples. We note that an additional 1.1\% of all binaries have $a_1 < 100$ pc and $10 < (a_2/a_1)_{\rm min} \leq 100$. Systems that nearly missed forming WTs ($1 < (a_2/a_1)_{\rm min} \leq 10$ while $a_1 > 100$ pc) are more common, representing 30\% of all FTs, or 3.8\% of all binaries. 

  \begin{figure*}
    \centering
    \begin{minipage}[b]{.48\textwidth}
    \includegraphics[width=3.5in]{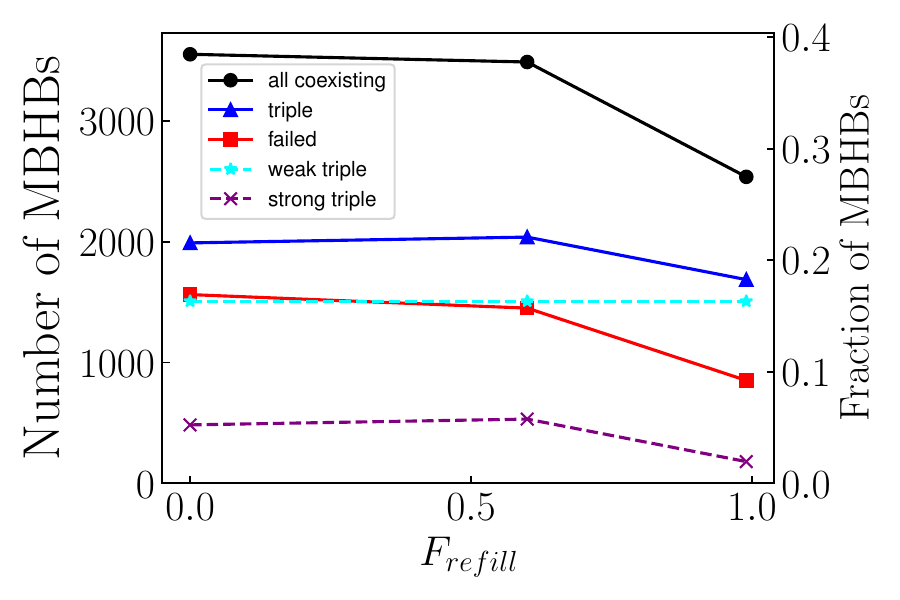}
    \end{minipage}\qquad
    \begin{minipage}[b]{.48\textwidth}
    \includegraphics[width=3.5in]{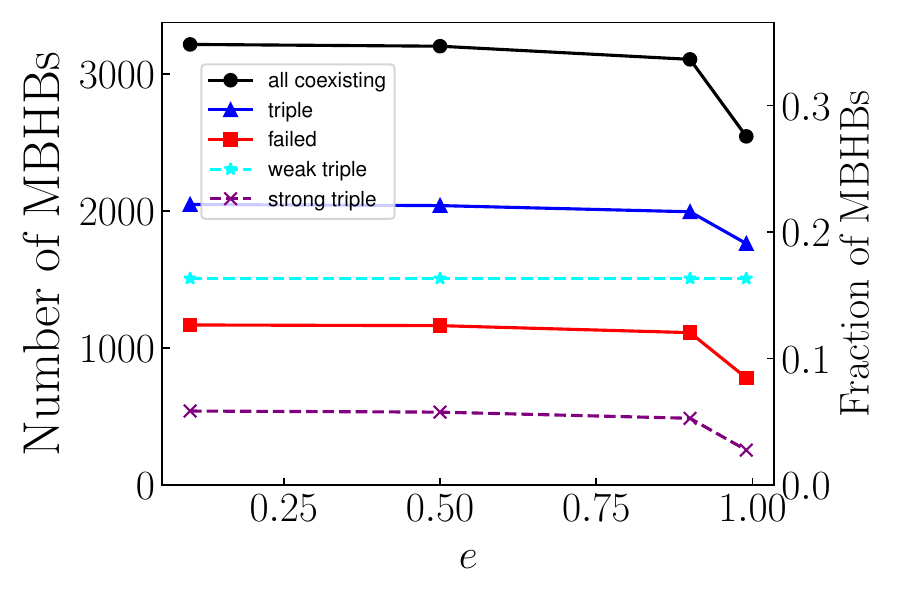}
    \end{minipage}
    \caption[Dependence of triple occurrence on LC refilling and eccentricity]{%\laura{\bf[I went ahead and switched the order of these plots to match the order of the discussion in the text. i also edited the caption a bit.]} 
    The dependence of the triple population on the LC refilling efficiency ($\mathcal{F_{\rm refill}}$, left panel) and binary eccentricity (right panel) is shown here. In both panels, all coexisting binaries are indicated in black; they are the superset of all other triple populations. Strong triples (dashed magenta) and weak triples (dashed cyan) constitute the total triple (solid blue) population. Failed triples (red) are the systems in which the second coexisting binary either never overtakes the first (i.e., the intruder MBH ``chases" the inner binary), or it fails to do so before $z=0$. Roughly 1/3 of all MBH binaries coexist at some point with another binary (i.e., their inspirals overlap in time), and most of these become triple systems. A few percent of all MBH binaries form strong triples at small separations. These results are remarkably insensitive to variation in binary inspiral model parameters.}
    \label{fig:model_dependence}
  \end{figure*}

  \begin{figure*}
    \centering
    \includegraphics[width=6.6in]{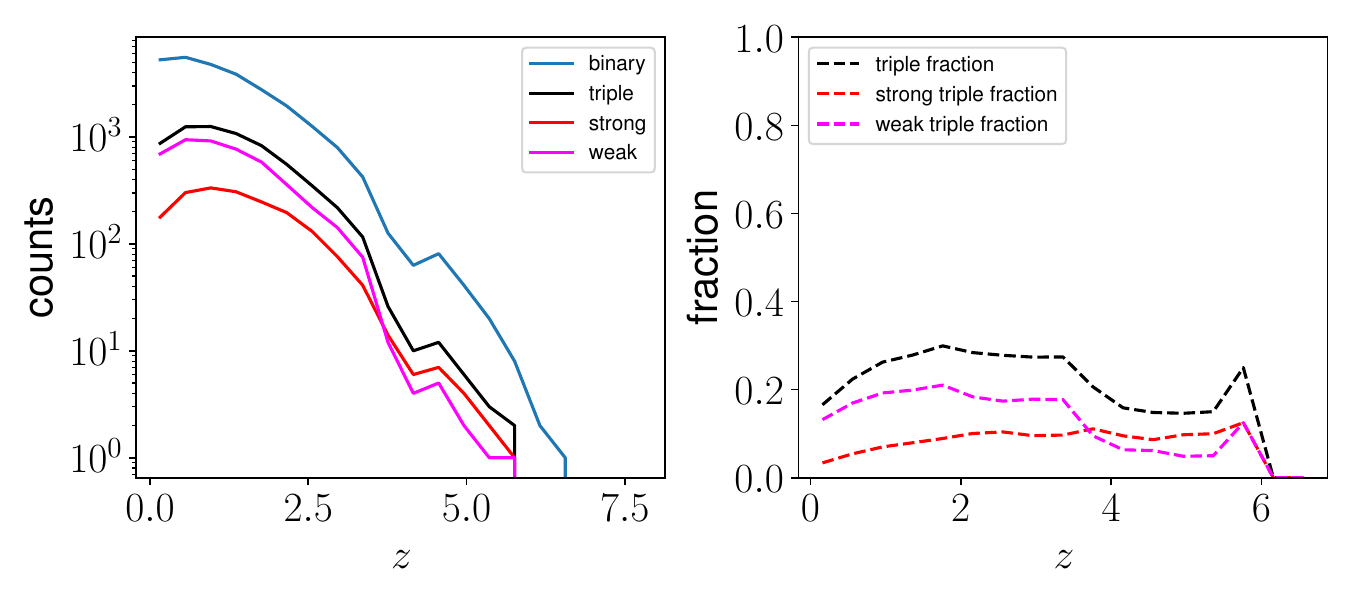}
    \caption[D]{The evolution of all binaries and various triple sub-populations are shown here as a function of redshift for the fiducial model. The left plot shows the total numbers and the right plot shows the triple populations as a fraction of the total number of binaries. From the left panel we can see that the triple population increases as the total binary population increases, as expected, and turns over at $z\lesssim 1$ as binary mergers begin to outweigh new binary formation. The fraction of triples remains fairly flat over time, $\sim 0.2 - 0.3$. Surprisingly, strong triples initially outnumber weak triples, but this trend reverses at $z \sim 3.7$. At redshift close to zero we see that the majority of triple systems are weak triples. }
    \label{fig:triples_zdist}
  \end{figure*}

\subsection{Dependence on Binary Inspiral Model Parameters}\label{model_dependence}
Here we explore the dependence of the triple MBH population characteristics on two key binary inspiral parameters: the stellar LC refilling efficiency $\mathcal{F_{\rm refill}}$ (as defined in Equation \ref{refill_fraction}) and binary eccentricity. Recall that we start the binary hardening process by assigning the same initial eccentricity to all binaries, after which the eccentricity evolves through the LC and GW stages of evolution according to Equations \ref{dedt_lc}-\ref{dadt_gw}. Note that the DF and CD phases of evolution are unaffected by variation in these two binary parameters.
  
Fig \ref{fig:model_dependence} shows how the various triple subsets depend on the LC refill rate and the initial binary eccentricity. These plots illustrate an important finding of our analysis: the number of triple systems does {\em not} depend very strongly on either $\mathcal{F_{\rm refill}}$ or eccentricity. For weak triple systems, this is essentially true by definition: at separations $> 100$ pc, binary hardening is not dominated by LC scattering or GW emission, so neither $\mathcal{F_{\rm refill}}$ or eccentricity has an impact on the occurrence of WTs. Less intuitive is the similar number of ST systems across a wide range in $\mathcal{F_{\rm refill}}$ and eccentricity values. 

We can understand the lack of dependence on LC refilling efficiency as a balance between two competing effects: moderate LC refilling can slightly enhance the triple population by bringing in third MBHs more quickly, but in the full loss cone regime, this is more than offset by the increased likelihood that the inner binary will merge before being overtaken by a third BH. Nonetheless, we find that the population of strong triples is remarkably robust---at a few percent of the total binary population---whether the LC is full, empty, or somewhere in between. 

When $\mathcal{F_{\rm refill}}$ is increased from zero (a steady-state, empty LC) to 0.6 (a partially full LC), the total number of coexisting binaries decreases slightly, because shorter inspiral timescales increase the likelihood that the first binary will merge before the second is formed. However, the triple fraction goes {\em up} very slightly (by half a percentage point, from 21.6\% to 22.1\%) over the same range in $\mathcal{F_{\rm refill}}$ values. This arises from systems in which the increased LC refilling efficiency has an outsized effect on the outer binary relative to the inner binary. In other words, some systems that are failed triples in an empty loss-cone model become successful triples when moderate LC refilling is assumed.

In contrast, when the LC is always full ($\mathcal{F_{\rm refill}}=1$), the faster binary hardening significantly reduces the chance of coexistence. While 38.5\% of all binaries coexist with another binary during their inspiral in the $\mathcal{F_{\rm refill}}=0$ model, only 27.5\% of binaries coexist in the full LC model. This reduces the fraction of binaries that form strong triple systems as well: 5.2\% and 5.8\% of all binaries form STs for $\mathcal{F_{\rm refill}}=0$ and $\mathcal{F_{\rm refill}}=0.6$, respectively, versus 1.9\% of all binaries that form STs for $\mathcal{F_{\rm refill}}=1$.
  
Figure \ref{fig:model_dependence} reveals a similarly weak dependence of triple occurrence on binary eccentricity. Again, the binary eccentricity is not expected to have any effect on weak triples in our model, but even for strong triples, the ST fraction varies by less than 1\% when the initial binary eccentricities are varied from $e=0.1$ to $e=0.9$. This reflects the fact that eccentricity influences binary hardening rates predominantly in the GW phase, which represents a small portion of the total inspiral time. Only at truly extreme initial eccentricities of $e=0.99$ do we see a noticeable drop in the occurrence of triple systems, owing to the strong dependence of GW-driven hardening on eccentricity (Equation \ref{dadt_gw}). At very high eccentricities, strong GW emission at pericenter can greatly reduce the merger timescale and drive the inner binary to merger before it is overtaken by an intruder MBH.

The redshift distribution of binaries, triples and their strong and weak sub-population are shown in Figure \ref{fig:triples_zdist} with the total numbers indicated on the left panel and the fractions on the right panel. The weak triples fraction increases over time at the expense of the strong triple population. The rich environment of early galaxies along with higher merger rates could contribute to the higher proportion of strong triples.  
  
Overall, we see that the triple population is not strongly dependent on the specific hardening model we choose. This stands in stark contrast to the $z=0$ population of un-merged binaries, which varies widely depending on the binary inspiral model parameters assumed \citep[most notably the LC refilling efficiency;][SB21]{2017MNRAS.464.3131K, 2017MNRAS.471.4508K}. The fraction of binaries that evolve to become triple systems, however, remains essentially unchanged across a wide range of model variations, except for very high eccentricities and full loss cones. Our models therefore predict that a robust $\sim$ few percent of all MBH binaries will form ST systems. Notably, these results suggest that the fraction of binaries that evolve to become triples is largely determined by the galaxy merger history, rather than the details of the astrophysical environment driving binary inspiral. This underscores the importance of using full cosmological hydrodynamics simulations instead of semi-analytic models to make these predictions, given that these methods have been shown to produce significantly different trends in the galaxy merger rates with redshift and stellar mass \citep[]{2015MNRAS.449...49R}.

\subsection{Host Galaxy Properties}\label{host_property}

\begin{figure}
    \centering
    \includegraphics[width=\columnwidth, trim=0.15in 0 0.14in 0, clip]{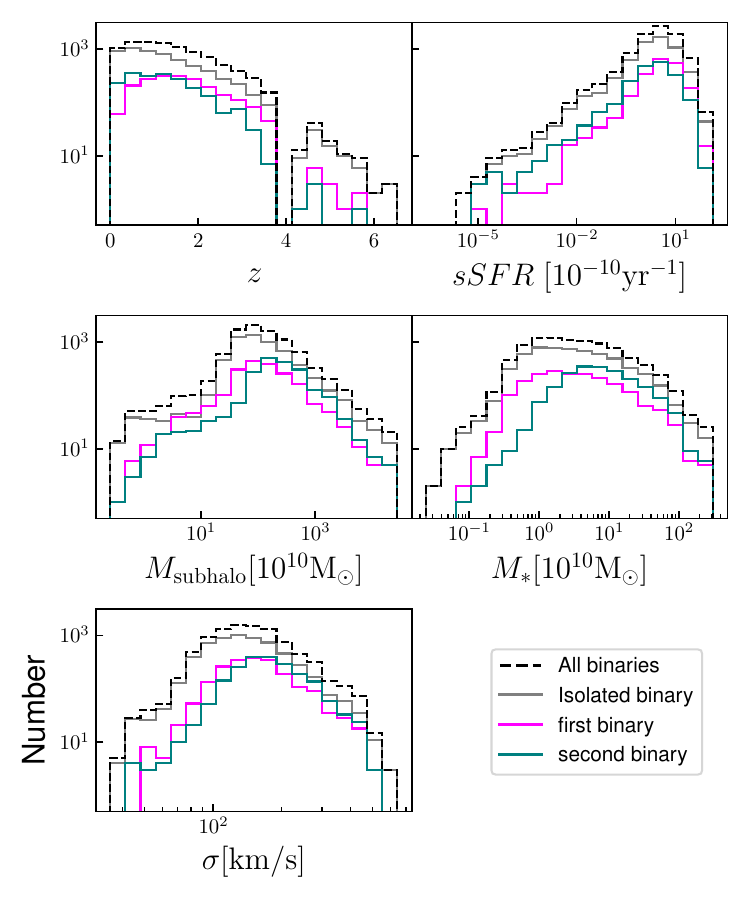}
    \caption[Galaxy parameters for the binary sub-populations]{Distributions of several properties of MBH binary host galaxies are shown: binary formation redshift (top left panel), sSFR (top right), total galaxy mass $M_{\rm subhalo}$ (middle left), stellar mass $M_*$ (middle right), and stellar velocity dispersion (bottom left). The black dashed lines show the histograms for the hosts of all 9234 MBH binaries (dashed line). The solid-line histograms show the sub-populations of isolated binaries that never coexist with another binary (grey), the first binary in each coexisting system (magenta), and the second binary in each system (cyan). Only modest differences are seen between these sub-populations, which reflect the fact that the hosts of first binaries, forming by definition at earlier times, have characteristics more typical of higher-redshift galaxies. Note that the discontinuity in the redshift corresponds to two corrupted Illustris snapshots at $z\sim4$ that are excluded from all analysis and do not affect our conclusions. 
    }
    \label{fig:env_all_binary}
\end{figure}

Clearly, we can expect that galaxies that host triple MBH systems will generally live in richer cosmic environments than those that do not undergo multiple mergers. Here we examine whether, in addition, the triple MBH host galaxies themselves may have distinctive features. 
In Figure  \ref{fig:env_all_binary} we look at key galaxy parameters for all  BH binaries  in Illustris, along with the sub-populations of first and second binaries in coexisting systems, as well as the isolated binaries that never coexist with another binary, which comprise 65\% of the binary population in our fiducial model. 

Figure \ref{fig:env_all_binary} shows the host properties for each of these sub-populations in the simulation snapshot immediately following the formation of the MBH binary. Overall, we see that the distributions of host galaxy properties look very similar for isolated binaries versus all binaries, as expected given their dominance of the population. There is, however, a noticeable difference between the first binaries and second binaries in coexisting systems. By definition, the inner binary (first binary) in a coexisting system forms at slightly higher redshift than the outer binary (second binary). This is reflected in the minor differences between their host galaxy properties, such that on average, the hosts of first binaries have lower stellar masses, lower stellar velocity dispersions, and higher specific star formation rates (sSFR, defined as the star formation rate divided by the stellar mass). 

\begin{figure*}
    \centering
    \includegraphics[width=0.9\textwidth, trim=0.3cm 5cm 0.25cm 3cm, clip]{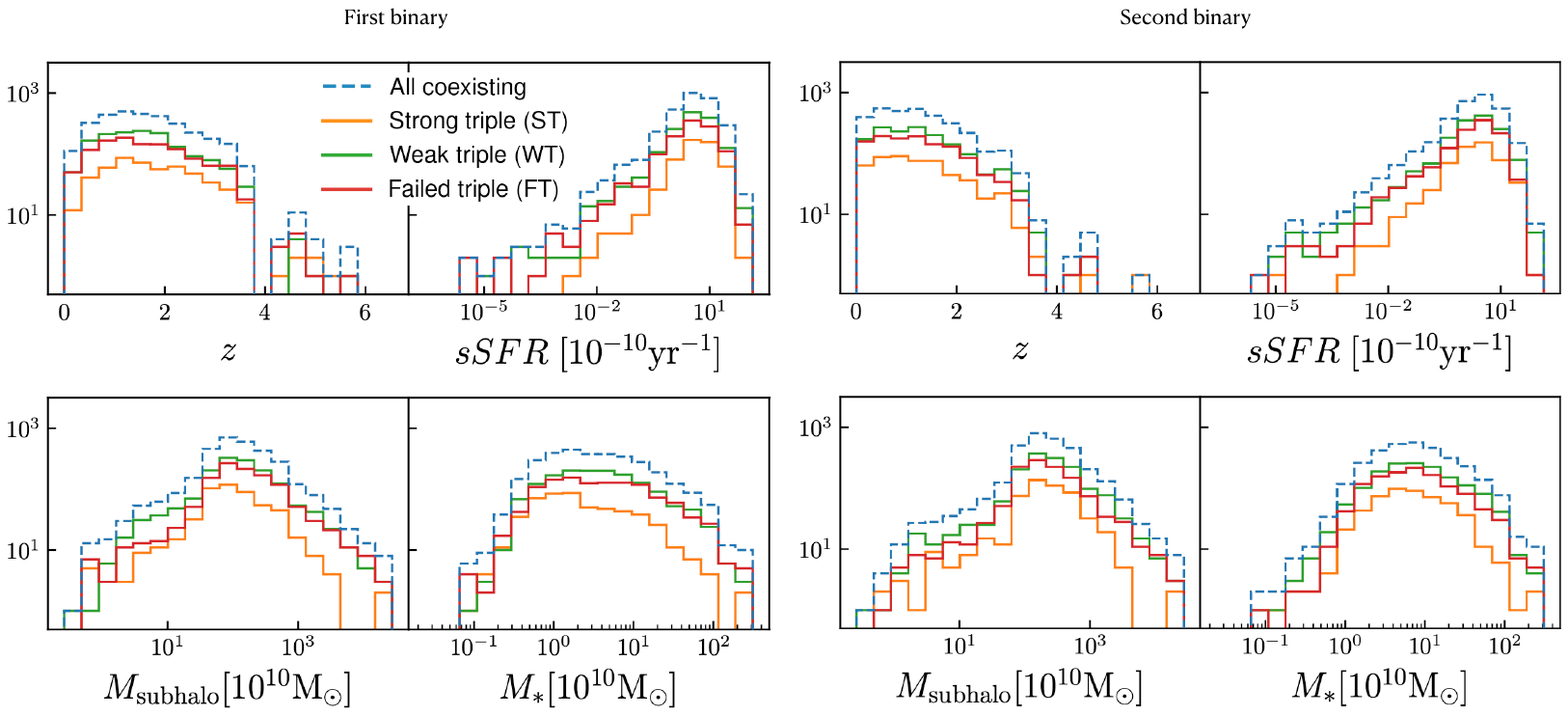}
    \caption[Environment parameters for the triple sub-populations]{Distributions of host galaxy properties are shown for the first binary (left) and second binary (right) in each coexisting binary system, in the simulation snapshot immediately after the first binary forms. Host properties are plotted in the same manner as in Figure \ref{fig:env_all_binary}. Here, the blue dashed lines show the host distributions for the each of the first and second binary in each coexisting system, while the solid-line histograms show the host distributions for each of the triple sub-populations: strong triples (orange), weak triples (green), and failed triples (red). While the distributions are fairly similar overall, the progenitors of strong triples form at slightly higher redshift in galaxies with slightly lower masses and higher sSFRs. Note that as mentioned before, the discontinuity in the redshift is due to 
    two corrupted Illustris snapshots at $z\sim4$.} 
    \label{fig:env_all_triple}
\end{figure*}

Figure \ref{fig:env_all_triple} shows the host properties for the first binaries (left) and the second binaries (right) in coexisting and triple systems. We see that the host properties are broadly similar across the sub-categories of STs, WTs, FTs, and all coexisting systems. There is a slight shift in the redshift distribution, as the second binaries form later in time, and the stellar mass of the second binary peaks at a higher value. Aside from that all the distributions for the sub-populations of WTs, STs, FTs, and all coexisting systems are qualitatively similar.  
  
Binaries that later become strong triples after being overtaken by an intruder do tend to form at slightly higher redshift. These ST progenitors also have somewhat lower masses and velocity dispersions and slightly higher sSFRs. Therefore, while none of these differences are large, the host properties that distinguish the first binaries in triple systems from other binaries are even more pronounced for the hosts of first binaries that go on to form strong triples. This can be understood in part by recalling that weak triples have somewhat lower median inner mass ratios than strong triples ($q_{\rm inner}\sim0.1$ versus $q_{\rm inner}\sim0.3$), and that nearly all of the triples with very low mass ratios ($q_{\rm inner}<0.01$) are weak triples. Galaxies at $z\gtrsim1-2$ will on average have lower masses (and thus higher minimum mass ratios), higher gas content (yielding higher SFRs and more efficient binary inspiral), and higher galaxy merger rates (such that an intruder is more likely to overtake the first binary before it merges). In addition, because strong triples typically have a long delay between the first binary formation and the triple formation, some lower-redshift binaries that might have become strong triples are likely to end up instead as failed triples, because the intruder is unable to overtake the first binary before $z=0$.

\section{Discussion}\label{discussion}

In this paper we have identified and characterized a population of candidate triple MBH systems in the Illustris cosmological simulation, using a new classification scheme that we have developed. We first assign sub-resolution binary inspiral timescales to each merging MBH pair from the simulation, using the binary inspiral models of \citet{2017MNRAS.464.3131K}. When a given MBH is involved in two successive binary inspirals that coexist in time, this system is identified as a possible triple MBH. If at any point in time the separation of the second binary becomes smaller than the separation of the first binary, this system is identified as a successful triple MBH. Because this analysis is based on MBH binary models that assume each binary evolves in isolation, these inspiral timescales do not incorporate the dynamical effects of a third, intruder MBH. We also do not account for any MBH mass growth that may occur during the binary inspiral \citep[cf.][]{2020MNRAS.498..537S}. Nonetheless, our findings predict that a non-negligible fraction of MBHs are involved in triple systems at some point in their evolution and that this population is relatively insensitive to assumed binary inspiral model parameters. Our work also highlights some key characteristics of this putative triple MBH population.

We find that more than a third (35\%) of all MBH binaries in Illustris do coexist with another binary, according to our fiducial inspiral model. Most of these (and a quarter of all Illustris MBH binaries) are in fact overtaken by another MBH at some point, and in nearly all such cases (22\% of all binaries), the MBH binary is overtaken before $z=0$. Our findings are consistent with \citet{2017MNRAS.464.3131K}, where they used a similar method to estimate that $\sim30$\% of all Illustris MBH binaries could form triple systems. Similar results were also obtained from the detailed semi-analytic models of \citet{2018MNRAS.477.3910B} and \citet{2019MNRAS.486.4044B}, which indicated that triple interactions could be an important channel for low-frequency GW sources observable with PTAs and LISA. 

Another key finding of this work is that the triple MBH population is bimodal. Triple systems can be clearly divided into two main sub-categories, which we refer to as ``strong triples" and ``weak triples," depending on the binary separation at which the first binary is overtaken by the intruder MBH (defined as the radius of triple formation, or $a_{\rm triple}$). The weak triples are defined as those that form at large separations, $a_{\rm triple}>100$ pc, with a median of $a_{\rm triple}\approx$ 990 pc, and these constitute 16\% of the total binary population. Strong triples ($a_{\rm triple}<100$ pc) form at a median separation of $a_{\rm triple}\approx$ 0.9 pc, and they constitute 6\% of the total binary population. This bimodality can be understood as the result of two different mechanisms that can lead to binary stalling: inefficient dynamical friction for inspiraling binaries at large separations, or inefficient stellar LC scattering at small separations. If indeed gaseous circumbinary disks do at times {\em increase} binary separations rather than drive efficient inspiral, this process may contribute to binary stalling at small scales as well. 

We find that strong triples typically require a lengthy delay of a few Gyr to evolve from the formation of the second binary to the point where the intruder MBH overtakes the first binary. In contrast, weak triples (which form at much larger binary separations) have a much shorter typical delay of a few $\times 10^7$ yr between second binary formation and triple formation. As a result, strong triples form at lower median redshifts than weak triples ($z=0.7$ versus $z=1.1$). Strong triples typically have a smaller total binary mass ($\rm M_{bin}$) and higher inner mass ratios as compared to the weak triples. The difference between the mass ratios for the outer binary is less significant.

Interestingly, we also find that the occurrence rate of triple systems is relatively independent of the binary inspiral parameters in the LC and GW phases. Specifically, we vary the initial binary eccentricity and the LC refilling rate, which by definition in our model do not affect binary evolution in the dynamical friction phase. This stage of binary evolution corresponds to the separations at which weak triples form. Thus, variations of these two parameters (binary eccentricity and LC refilling rate) do not affect weak triples at all. Somewhat surprisingly, though, we find that the incidence of {\em strong} triples is also relatively insensitive to these binary model parameters. In essence, this means that more efficient binary inspiral (due to higher eccentricity or faster LC refilling) shortens merger timescales and triple formation timescales by roughly similar amounts. We do, however, see a decrease in triple formation when stellar loss cones are assumed to be always full: $\sim 1.9\%$ of all binaries form strong triples in this case, as opposed to $\sim 5 - 6\%$ for empty loss cones and partially refilled loss cones. This indicates that when stellar hardening in the LC phase is very efficient, binaries do tend to merge faster than triples can form. But given the wide range in binary inspiral timescales between full and empty loss cones \citep{2017MNRAS.464.3131K}, the influence on triple occurrence is relatively modest. 

Variations in binary eccentricity (which impact evolution in the LC and GW phases) have even less influence on the triple population; only for the extreme model with $e=0.99$ do we see an appreciable drop in the formation rate of strong triples. Note that we do not consider any eccentricity evolution in the dynamical friction or gas-driven binary inspiral phases. The circumbinary disk phase in particular depends on the MBH accretion rates (SB21), so a variation in the accretion model could also change these results. These questions will be explored further in future work.

Regardless, because we find that a consistent $\sim$ few percent of all MBH binaries are involved in a strong triple interactions, we can conclude that the occurrence of triple MBH systems is closely tied to the rate of MBH binary formation and by extension to the galaxy merger rate. This highlights the importance of accurate measurements of the galaxy merger rate as an essential foundation for constraining MBH dynamics and gravitational wave source populations.

One of the many complexities of triple MBH dynamics is that the galaxy nucleus containing the first binary may be significantly altered by the subsequent galaxy merger that introduces the intruder MBH. Merger-driven gravitational instabilities can perturb the existing stellar distribution and can also drive cold gas inflows, which may trigger new nuclear star formation. Thus, if the first binary's inspiral had stalled due to an empty stellar loss cone, a subsequent galaxy merger would provide a possible means for partially refilling the loss cone on roughly a dynamical timescale. This could in theory shorten the merger timescale of the first MBH binary, {\em before} the intruder MBH reaches the nucleus to undergo a strong triple interaction. In some cases, this might even mean that the first MBH binary could merge before a triple system was able to form. As we have noted throughout, a full treatment of triple MBH dynamics is beyond the scope of this work. However, our finding that LC refilling has only a modest impact on the strong triple population can be used to constrain the influence of galaxy merger dynamics on triple formation.

The most extreme example of the above scenario would be one in which a stalled binary in an initially empty loss cone is quickly refilled by galaxy merger dynamics (i.e., perturbations to the stellar nucleus, or formation of a new stellar cusp). We can estimate the impact of such a scenario on the triple population by comparing the two extremes of our LC refilling model variation: $\mathcal{F}_{\rm refill} = 0$ and $\mathcal{F}_{\rm refill} = 1$. Figure \ref{fig:hardening_scatter} shows that the proportion of all binaries that form strong triples drops from $\sim 5\%$ to $\sim 2\%$ when loss cones are assumed to be completely full instead of completely empty. Therefore, the impact of galaxy merger dynamics on LC refilling (and thus MBH merger timescales) would have to be smaller than this. In other words, our results indicate that merger-driven LC refilling could reduce but not decimate the triple MBH population.

We additionally note that, because this study models the evolution of each binary in isolation, some systems that  form a failed triple can eventually develop into either weak triples or strong triples. And some systems that initially form as weak triples could potentially develop into strong triples at a later time. This can happen if the $(t,\;a(t))$ components of one binary cross that of the other binary multiple times. For example, if the second binary initially has a higher rate of hardening and the intruder MBH overtakes the first binary at $a(t) \sim 1$ kpc, a weak triple system is formed. If this second binary later stalls at $\sim$ pc separations, the first binary could in theory overtake it again, leading to a strong triple interaction. Around $\sim 19\%$ of the triple population experiences more than one crossing. Of course, in reality the evolution of the inner and outer binaries is not isolated, and their dynamics are more complicated.

Analysis of host galaxy properties reveals some mild distinctions between binary and triple MBH hosts. The first binary in each triple system forms at higher redshift than the second, by definition. Accordingly, the median total and stellar host masses are $\sim 2-3$ times lower when the first binary forms than when the second binary forms. The hosts of the first binaries also have $\sim 2$ times higher sSFR. Among the first binaries in each triple, those that eventually undergo strong interactions with the intruder MBH form at slightly higher redshifts than those that become weak triples. The higher major merger rates of high-redshift galaxies are favorable for the formation of triple systems (note that a significant tail of strong triples form at $z<1$). These hosts of strong triples therefore tend to have lower masses and higher sSFRs than the hosts of weak triples. None of these trends in host galaxy properties are dramatic, however, and their distributions all have substantial overlap. 

While the weak triple systems are too widely separated to allow interactions between MBHs, they have an advantage in terms of observability, because their constituent MBHs could be spatially resolved. If more than one of these MBHs is actively accreting, they may appear as a dual or triple AGN system. Such objects are of great interest, in part because they provide information about the early stages of MBH binary formation and inspiral. Moreover, multiple-AGN systems can provide clear examples of AGN fueling induced by galaxy mergers \citep[]{pfeifle19b}. The role of mergers in AGN fueling is actively debated, with some studies finding a clear merger-AGN connection, and others finding that mergers are not an important channel for triggering AGN \citep[e.g.,][]{ellison11, cisternas11, kocevski15, villforth17, koss18}. Triple AGN systems, by definition, form in hosts with an active recent merger history; thus, further observations of such systems would offer a unique window into the merger-AGN connection.

Some of the weak triple systems we identify in Illustris consist of a massive primary MBH, a much less massive secondary MBH, and an intruder MBH that may or may not be massive. In such cases, the low-mass MBH(s) are likely to have long inspiral timescales and may be difficult to detect unless they are accreting at a substantial fraction of the Eddington rate. Future observational constraints on the prevalence of dual and triple AGN systems, including those with small MBH mass ratios, will enable further studies of the weak triple population. In addition to revealing key information about the nature of AGN fueling in galaxy mergers, discoveries of these systems will also provide implicit constraints on the strong triple population.

The strong triples can lead to the rapid merger of any of the binary members of the triple, which may greatly reduce the merger time if the binary inspiral had previously been stalled. In a study of the parameter space of triple MBH interactions, \citet{2018MNRAS.477.3910B} found that $\sim$ 20 - 30\% of binaries that would not otherwise have merged within a Hubble time were driven to merger by triple interactions. In such cases where the coalescence happens before $z=0$, these triple interactions can increase the MBH merger rates detectable with LISA. Likewise, MBH binaries stalled at $\gg$ mpc separations will not produce GWs detectable with PTAs, but triple MBH interactions can rapidly harden these binaries and bring them into the GW regime, where they could contribute to the stochastic GW background or even be detectable as continuous-wave GW sources at $\sim$ nHz frequencies. Meanwhile, in a hierarchical 3-body system, the intruder MBH can trigger K-L oscillations, which can rapidly bring the inner MBHs to the GW dominated regime. This can result in an MBH merger detectable with LISA, possibly preceded by bursts of GW emission at frequencies within the LISA band.

Strong, chaotic three-body interactions can also eject the lightest MBH from the system while further hardening the more massive pair. Thus, in addition to their impact on merger rates, triple interactions can create a population of offset MBHs. Note that a galaxy nucleus still contains one or more MBHs after a slingshot recoil, in contrast to a GW recoil that ejects the merged MBH and leaves behind an empty galactic nucleus. If the remaining MBH pair merges and experiences a large GW recoil kick, however, this could lead to multiple offset AGN {\em and} an empty nucleus. Further theoretical studies are needed to understand the relative importance of slingshot versus GW recoil for offset MBH populations and for MBH-galaxy co-evolution.
        
Our main results can be summarized as follows:

\begin{itemize}
    \item {\bf Significant triple population: }35\% of all binary inspirals coexist when evolved in isolation, and 22\% of all binaries form triples, defined as the second binary overtaking the first.
    
    \item {\bf Strong and weak triples sub-populations: }The true triple population can be categorized into  two distinct sub-populations of strongly interacting ($a_{\rm triple}<100\rm pc$) and weakly interacting ($a_{\rm triple}>100\rm pc$) triples based on the radius at which the triple forms. In our fiducial model, the strong and weak triples comprise 6\% and 16\% of the total binary population, respectively.
    
    \item {\bf Lack of dependence on loss cone refilling or binary eccentricity: } The binary loss-cone refilling parameter has a fairly minor effect on both weak and strong triple populations. Only for very high values of ($F_{refill} \sim 1$) do we see a noticeable drop in the triple population, which is mostly caused by first binaries merging before they are overtaken by an intruder MBH. Similarly, the triple population and all of its subcategories are remarkably invariant to the initial binary eccentricity. We notice a drop in the triple population only for highly eccentric binaries ($e\sim 1$), where very rapid binary evolution drives MBHs to merger before the intruder MBH can intervene.
    
    \item {\bf Binary and triple MBH host properties: } The first binaries and the second binaries in triple systems have slightly different host properties. Hosts of the first binary have lower stellar masses and stellar velocity dispersion and slightly higher sSFR, consistent with their slightly higher formation redshifts. Weak and failed triples have nearly identical distribution of host properties for both the initial binary host and the host for the second merger (which creates the triple system). For strong triples, on the other hand, the first binary host has slightly smaller subhalo masses and stellar masses. This can be understood as a combination of somewhat larger mass MBH ratios and longer delay times between first binary formation and strong triple formation, which favors first binaries that form at slightly higher redshifts. 

\end{itemize}

\section*{Acknowledgements}

This work made use of the \texttt{PYTHON} \citep{python} programming language along with Jupyter notebooks (Kluyver et al. 2016). Visualizations are made with MatplotLib library \citep{Hunter:2007}. Numpy was used for the analysis \citep{harris2020array}. L.B. acknowledges support from NSF grant AST-1909933, NASA grant 80NSSC20K0502, and from the Research Corporation for Science Advancement via Cottrell Scholar Award 27553.

\section*{Data Availability}
The underlying data for the Illustris cosmological simulations are publicly  available \citep{2015A&C....13...12N}. The data underlying the sub-resolution analysis and MBH binary evolution  will be shared on reasonable request to the corresponding author.

\bibliographystyle{mnras_tex_edited}
\bibliography{main} %

\label{lastpage}
\end{document}